\documentclass[amsbook,amsmath,amssymb,byrevtex,dvips,showpacs]{revtex4}
\usepackage{graphicx,amssymb,latexsym,amsmath,amsbsy}

\newcommand{\etal}{{\it et al.,\;}}
\newcommand{\beq}{\begin{equation}}
\newcommand{\eeq}{\end{equation}}
\newcommand{\bea}{\begin{eqnarray}}
\newcommand{\eea}{\end{eqnarray}}

\newcommand{\benn}{\begin{displaymath}}
\newcommand{\eenn}{\end{displaymath}}

\begin{document}

\title{\large The Long Journey from {\it Ab Initio} Calculations to Density Functional Theory for 
Nuclear Large Amplitude Collective Motion }

\author{ Aurel Bulgac  }
\affiliation{Department of Physics, University of
Washington, Seattle, WA 98195--1560, USA}

\begin{abstract}
 
At present there are two vastly different {\it ab initio} approaches to the description of the the many-body dynamics: the Density Functional Theory (DFT) and the functional integral (path integral) approaches. On one hand, if implemented exactly, the DFT approach can allow in principle the exact evaluation of arbitrary one-body observable. However, when applied to Large Amplitude Collective Motion (LACM) this approach needs to be extended in order to accommodate the phenomenon of surface-hoping, when adiabaticity is strongly violated and the description of a system using a single (generalized) Slater determinant is not valid anymore. The functional integral approach on the other hand does not appear to have such restrictions, but its implementation does not appear to be straightforward endeavor. However, within a functional integral approach one seems to be able to evaluate in principle any kind of observables, such as  the fragment mass and energy distributions in nuclear fission. These two radically approaches can likely be brought brought together by formulating a stochastic time-dependent DFT approach to many-body dynamics.  
 
\end{abstract}

\date{\today}


\maketitle

\numberwithin{equation}{section} 

\section{ Difficulties in implementing a theoretical framework for the nuclear Large Amplitude Collective Motion }

We know a great deal by now about interactions between nucleons, although one might rightly argue that the link to quarks and gluons through the more fundamental theory Quantum Chromodynamics has not yet been fully established. It will still be a long time until this goal will be reached with the needed degree of accuracy and sophistication. In this respect the nuclear many-body theory is at a great disadvantage when compared to condensed matter and atomic theories or chemistry, where the dominant role of the Coulomb interaction and the link to Quantum Electrodynamics were clarified a long time ago. There is a general consensus by now however, that in order to describe nuclear many-body systems one probably can get away with the use of one set or another of reasonably well established two- and three-nucleon potentials (if only one would manage to use them in a convincing implementation) and that a non-relativistic approach is likely sufficient for now.  The best example we have so far that such an {\it ab initio} approach can be brought to fruition is the Green Function Monte-Carlo (GFMC) approach implemented so successfully by V.R. Pandharipande and his many students and collaborators J. Carlson (truly the author of GFMC), S.C. Pieper, R. Schiavilla, K.E. Schmidt, R.B. Wiringa,  and many others \cite{gfmc}. One of the most frustrating limitations of the GFMC approach is the relatively low particle number this method can deal with. Presently one cannot envision the study of nuclei heavier than oxygen within a GFMC framework. The Coupled-Cluster (CC) method \cite{cc} looks very promising in this respect, as it might be applied to medium size closed shell nuclei in the near future.  The No-Core-Shell Model (NCSM) \cite{ncsm} will likely be successfully applied to light and medium size nuclei in the foreseeable future with a reasonable degree of accuracy. 

Heavy-nuclei seem out of reach from a direct frontal {\it ab initio} attack and the only alternative appears to be an indirect one, namely through the use of the Density Functional Theory (DFT) \cite{HK}, and only if one would be able to generate a sufficiently accurate nuclear energy density functional. During the last few years many theorists became rather optimistic in this respect and the goal could prove to be within reach \cite{UNEDF}. The strategy proclaimed by the UNEDF group \cite{UNEDF} can be rather briefly formulated as follows: assuming that a universal nuclear density functional exists and that it can be written down (whether in a local or non-local form is still somewhat a matter of debate) one can verify it by describing light and hopefully medium nuclei, many properties of which could be described within one or another {\it ab initio} approach (GFMC, CC, NCSM). In this way one would test one theoretical approach against the other. Such an {\it ab initio} inspired energy density functional could be put to further test and validated by computing a range of nuclear properties throughout the whole nuclear table and insuring that a desired degree of accuracy has been reached, assuming also that the magnitude of the theoretical errors could be also estimated. 

Within a DFT framework  one can calculate only the total binding energy and one-body properties. With the appropriate extension to time-dependent phenomena, when a Time-Dependent DFT (TDDFT) approach is implemented \cite{td-dft,tddft}, one can address the properties of a number of excited states as well, basically those states which can be excited by a weak external probe and which can be described within the linear response regime and even a rather wide class of nuclear reactions. While reaching this goals and putting on a firm basis the description of a large number of nuclear properties will be a truly heroic endeavor, there are a number of extremely important nuclear phenomena and properties, which clearly fall outside a DFT-like description. Generically, any excitation which is not of small amplitude, and thus one which is not amenable to a description within the linear response regime, is of this type. Perhaps the most prominent example of nuclear Large Amplitude Collective Motion (LACM) is the nuclear fission, a phenomenon discovered seven decades ago and which is still waiting for its full microscopic description, and many would claim even its full understanding still.   

While proceeding along this road one should keep in mind however that a number of theoretical errors are introduced from the start, by assuming that the many-body dynamics can be described with a non-relativistic Hamiltonian, with instantaneous interactions, and also within the framework of quantum mechanics and not within a quantum field theory framework. There exist so far no reliable estimates of such corrections to the binding energy of $^{208}${\it Pb} for example, and no one can claim that these corrections are smaller or bigger than 1 MeV even. Mostly likely the latter is true. One can hardly make the claim that such effects are {\it effectively} included, either through the choice of the parameters or through the choice of the parameterization form. 

There is a wide consensus at this time in the community that the role of the three nucleon forces are crucial in nuclei \cite{gfmc,eft} and that a (the) major contribution arises from a virtual excitation of a $\Delta$-isobar in a three nucleon system.  A major assumption is also made that in a large nucleus the form of the NN and NNN interactions are not modified by the presence of the other nucleons, and implicitly also that the properties of nucleons are not modified by the medium. While there is clear evidence from the EMC-effect that that is not the case \cite{emc}, we lack any reliable estimates of how big the corrections to the binding energy of $^{208}${\it Pb} would be, and by how much its properties would be altered by an explicit treatment of $\Delta$-isobars in the nuclear medium and by the modification of the nucleon properties in nuclear medium.  Nobody can state that these corrections are for example either smaller or larger than 1 MeV for example. One cannot make the claim either that such effects are negligible, if the binding energies of nuclei is to be calculated within a specific phenomenological approach with an error smaller than 1 MeV, which is the state of the art in the case of nuclear mass phenomenological fits 
\cite{pearson,moller}. 

No phenomenological approach used so far in nuclear physics has ever been proven to be an exact theory, with or without a set of fitting parameters. Many argue that in order to be able to describe accurately heavy nuclei for example, one has to perform a fine tuning of various parameters of an energy density functional. By performing such a fine tuning however one clearly {\em sweeps under the rug} lots of physics. So far we simply do not know if a specific (semi-)phenomenological approach does indeed correspond to an accurate solution of the many-body non-relativistic Schr\"{o}dinger equation for a medium or heavy nucleus and we have no idea how precise is in principle the description of nuclei within a non-relativistic Schr\"{o}dinger equation approach to nuclei. In other words, so far we lack even an estimate of the order of magnitude of the theoretical errors inherited by adopting one or another theoretical framework. Consequently, any apparent agreement achieved between the description of nuclei within a specific (semi-)phenomenological approach and data {\em  can not} be used as an argument to make any inferences about the magnitude of other unaccounted effects.  

The main ideas behind the currently accepted theoretical framework to LACM go back more than half a century, to Hill and Wheeler \cite{hw,fsea}, and in its most modern formulation it comes under the name of Adiabatic Time-Dependent Hartree-Fock (ATDHF) theory \cite{rs}. The nuclear many-body wave function $\Psi$ is represented as
\beq
\Psi(x_1,...x_A) = 
\int \Pi_{k=1}^N dq_k\phi(q_1,...q_N)\Phi(x_1,...x_A|q_1,...q_N), 
\label{eq:coll_wf}
\eeq
where $x_1,...x_A$ are the (spatial and spin-isospin) nucleon coordinates and $q_1,...q_N$ are the collective variables. The wave function $\Phi(x_1,...x_A|q_1,...q_N)$ is chosen as a generalized Slater determinant (GSD), parameterized by the values of the collective variables, which in selfconsistent treatments are required to satisfy the constraints
\beq
q_l= \int \Pi_{k=1}^A dx_k \Phi^*(x_1,...x_A|q_1,...q_N)\hat{q}_l\Phi(x_1,...x_A|q_1,...q_N),
\quad \quad l=1,...,N, \label{eq:coll_op}
\eeq
and where $\hat{q}_l$ is an one-body operator (e.g. a particular quadrupole moment).
Once a set of collective variables has been chosen, either guided by the magical physical intuition of a given set of authors or by some other more formal set of rules \cite{abe}, one has to determine for each particular value of each collective coordinate the optimal GSD, which will minimize the total energy of a nuclear many-body Hamiltonian. This goal is typically achieved by solving a constrained self-consistent meanfield problem. 

All GSDs form a complete set of orthogonal many-body wave functions, and in principle the expansion in Eq. (\ref{eq:coll_wf}) has the potential to be exact. However, the idea put forward by Hill and Wheeler \cite{hw,fsea} impels us to use a much smaller set of such GSDs, which hopefully parameterize sufficiently accurately the sequence of nuclear shapes followed by a nucleus during the fissioning process for example. If one makes a {\em physically inspired educated guess} and an optimal set of GSDs has been determined, the next step is to determine the collective wave function $\phi(q_1,...q_N)$. The ensuing equation satisfied by the collective wave function is known as the Hill-Wheeler equation \cite{hw} and some other times it comes under the name of the Generator Coordinate Method \cite{rs}. In practice this approach often appears to lead to a surprisingly accurate description of a large number of nuclear data \cite{delaroche} and this success seems to lend support to the hope that an accurate description of many nuclear properties within a DFT-like approach will eventually be reached. While physically extremely appealing and intuitive, and also apparently backed by the agreement with data, this theoretical approach is plagued by a long list of serious drawbacks (some mentioned above, others to be discussed below), which have never really been seriously addressed. A serious drawback of this approach is the lack of a reliable measure of the theoretical error. This restricted representation of the many-body wave function could be exact if the many-body dynamics would be exactly separable by introducing a suitable set of intrinsic and  the collective coordinates $q_k$ \cite{abe}. Since this is never the case, the representation (\ref{eq:coll_wf}) is of an unquantifiable quality. The number and even the character of the {\em relevant} collective coordinates is not known and varies from author to author and often in the case of the same author from work to work, and it is essentially determined by the current computational capabilities of a given set of authors at a given moment in time, see Refs. \cite{delaroche,kenichi,pmoller,witek} and earlier references quoted therein.  Apart from the fact that these collective variables are not optimized, as discussed for example in Ref. \cite{abe}, in the sense that they are not optimally decoupled from the intrinsic degrees of freedom, the set of GSDs generated in this manner is also not complete and, moreover, there is a large degree of redundancy \cite{flocard} (as one can easily establish by computing the spectrum of the norm matrix). If collective degrees of freedom would be indeed well decoupled from the intrinsic motion, there would be no need to even discuss whether one should perform calculations with variation before or after the projection for example.

Even if one were to admit that the choice of the collective operators made by various authors is sensible, there are clear inconsistencies in using commonly introduced constraining operators. For example, if one uses the quadrupole moment $2z^2-x^2-y^2$ as a constraining operator and if one were to treat it mathematically exactly, this constraint would introduce unphysical modifications of the GSD, which are practically impossible to quantify. A constraining field $\lambda_{20}(2z^2-x^2-y^2)$ plays the role of a one-body potential which tends to $+\infty$ in some spatial directions and to $-\infty$ in others. Clearly single-particle wave functions in such a one-body field have a totally wrong spatial asymptotic behavior. If one would try to increase the size of the single-particle basis set used to diagonalize the self-consistent equations, the size of the errors introduced into the description would obviously increase. At the same time by decreasing the size of the single-particle basis one increases the size of the error as well, but for different reasons. One can then hope that there is an optimal basis set size for which the incurred error is minimal.  But even when this incurred error is minimal {\em  the size of the error is not known} and there are no known ways to even estimate it. No one knows which is the optimal size of the single-particle basis set, how this size varies with the proton and neutron numbers, how this size depends on deformation. It is not clear even if the size of the optimal single-particle basis set (when the incurred error is minimal) is similar for various constraining operators. {\em Thus it is absolutely unclear whether one can describe with any quantifiable accuracy the nuclear dynamics in the case of several collective variables within the commonly accepted parameterizations of the nuclear shape or otherwise. }

The only constructive procedure that I am aware of, which in principle could determine the relevant set of collective coordinates and also generate some measure of the accuracy of the representation (\ref{eq:coll_wf}) was developed by Klein and collaborators \cite{abe}. However, this approach is extremely difficult to implement in practice and the mathematical structure of the approach is not fully understood. In this approach the very character of the constraints depends on the collective variables themselves, and they have to be determined concurrently with the optimal GSDs. The collective inertia tensor is also determined concurrently along with a rather well defined measure of the quality  of the collective variables, but not of the size of the incurred error. If the quality of the collective variables is not satisfactory, one is required to increase the size of the collective space. It suffices to say that after a heroic  theoretical effort over several decades, only recently some relatively simple cases have been considered within a very simplified nuclear structure model \cite{kenichi}. The progress in this direction does not appear to be accelerating in the near future, and the prospects for success do not appear to be very encouraging. 

What makes matters even worse are a few additional aspects of an ATDHF-like approach. In most implementations of this scheme the collective coordinates are basically introduced based on {\em the magical physical intuition} of the authors. Usually there is no objective and unambiguous measure of the ``goodness'' of an assumed set of collective variables, neither of their form nor of their number. The typical set of collective coordinates considered in literature are the quadrupole deformations and sometimes a few higher multipoles, but even for the description of the same phenomenon various authors could not agree upon a unique, or even minimal (by some unknown yet measure) set of collective coordinates \cite{moller,witek}. In these approaches  the decoupling of the collective variables and optimization of the GSDs as required by theory \cite{abe} is never considered and thus their quality is simply unknown. One can safely state that as of now nobody knows how many collective coordinates are necessary in order to describe for example the fission dynamics of heavy nuclei with a predetermined and satisfactory accuracy, within the theoretical framework based on the representation  (\ref{eq:coll_wf}) of the total wave function. Success is typically declared when  agreement is achieved  with a selected subset of experimental data (by introducing and varying a relatively large number of parameters), and it is not determined intrinsically by the theoretical approach. When one notices differences between {\em theory} and experiment, one has the choice to claim that either one needs to increase the numerical accuracy of the theory, but at the same time one can also claim that some unaccounted physical phenomena could be responsible as well for that disagreement. 

Apart from the fact that there are more basis states than needed on physical grounds \cite{flocard}, it is basically impractical to deal with collective spaces of large dimensions. According to Ref. \cite{pmoller} one needs {\underline {at least}} five different collective variables and in their calculations these authors needed about five million different nuclear shape configurations. Only generating this number of constrained meanfield calculations requires basically a thousand times more computing time than for generating the entire nuclear mass table in a selfconsistent meanfield approach. It is obvious that even with significantly increased computer power envisioned during the next decade such an approach becomes unmanageable. In order to determine the collective wave function $\phi(q_1,...q_N)$ one needs to solve the integral Hill-Wheeler equation, the size of which becomes quickly extremely large for large dimensional collective spaces. For this reason most practitioners choose instead to reduce the Hill-Wheller equation to a differential equation, using the Gaussian Overlap Approximation \cite{rs}, the accuracy of which however has never been quantified. 

All in all, the method based on the representation of the nuclear many-body function (\ref{eq:coll_wf}) is not a controlled approximation, as it is not clear how big the theoretical error incurred really is. The usual approach chosen in practice, to {\em fine tune} the parameters of the Hamiltonian, or to increase the size of either the single-particle basis set or of the number of collective variables, are clearly not defensible arguments on theoretical grounds. One cannot achieve a faithful description of Nature when success is determined by more accurate fitting of the parameters of a clearly approximate approach, based on solving the non-relativistic Schr\"{o}dinger equation (which is an approximation to Nature from the start) and improve the agreement with experiment, when one has no idea about the theoretical errors incurred and when the approach is unfalsifiable by construction.  Moreover, one cannot use the current approach to test for physics beyond the initial assumptions. (For example, a phenomenon similar to the role of general relativity corrections in the precession of the perihelion of Mercury is clearly outside its scope of applicability.)

Another theoretical inconsistency in how the emerging Hill-Wheler/GCM equations are solved was never really discussed so far. Typically, in the case of spatial deformation one reduces the Hill-Wheller/GCM integral equations to a differential form, using as a rule the Gaussian Overlap Approximation or related approximations \cite{rs,delaroche}. However, when treating other collective degrees of freedom, such as particle number, isospin and/or angular momentum, typically one uses the integral formulation of the theory, as the collective wave function has a simple form and the theory is equivalent to merely a projection on the correct quantum numbers. In this case non-diagonal matrix elements need to be evaluated \cite{duguet} and rather arbitrary restrictions have to be imposed in order to avoid mathematical difficulties. There exist no defensible theoretical arguments why different collective degrees of freedom should be treated using non-equivalent theoretical schemes with different degrees of accuracy. 

There is a rather subtle argument against a DFT based approach to LACM, which as far as I know, has never been considered in the nuclear physics literature. Related issues have been addressed for quite some time in the description of the nuclear motion in molecules \cite{mdft}. While the description of the intrinsic motion is based on a density matrix approach, the collective dynamics is described by wave functions. It is not obvious that such a mixed approach makes any sense theoretically, and it has never been derived in any fashion, but it has been merely postulated implicitly. 

\section{ Possible resolution of these difficulties }

In order to evade the vicious circle described above one could proceed in a slightly different manner. There will be a price, and for some people this price may appear too high, as agreement between {\em theory} and experiment (which in my opinion in this case is over-rated) would naturally be worse at the beginning. But we will likely be able to achieve a deeper understanding, and that is by far a more valuable goal to strive for. Instead of trying to improve the quality of the description of a specific approach by allowing for the fine tuning of some parameters, one should instead fix such parameters before hand in a sensible manner and try to produce within a theoretically convincing approach the properties of medium and heavy nuclei with clearly defined and quantifiable errors, both theoretically and numerically. Any inconsistencies between theory and data which lie outside the error bars could then be attributed to phenomena not accounted for, such as relativistic effects, charge symmetry breaking of nuclear forces, modification of nucleon properties in the medium, excitation of other degrees of freedom, such as $\Delta$-isobars, and the potential modification of their properties in the medium, see for example Refs. \cite{brown,mosel}.  

The natural question arises, whether any sensible theoretical approach could be put forward in order to describe nuclear LACM in particular. I shall describe here an approach to LACM which brings together two rather different many-body techniques, each one of them in principle exact, but at the same time each one of them suffering from the fact that an exact implementation is more a matter of desire rather than fact, and each one of them still displaying a number of difficult unresolved aspects. 

\subsection{Real-Time Functional Integral Approach and some of its limitations}

The first approach is based on the real time functional integral treatment of the many-body problem \cite{qmps}. It can be easily shown that at least in principle one can represent the evolution operator for a  many-body system governed by a Hamiltonian $\hat{H}$ from an initial time $t_i$ to a final time $t_f$ as follows ($\hbar=1$ in this work):
\bea
\exp[-i\hat{H}(t_f-t_i)]&=& 
\int \Pi_{k=1}^{N_t}\Pi_{\alpha\beta}\quad  d\sigma_{\alpha\beta}(k) 
\exp\left [ \frac{i\Delta t}{2}\sum_{k=1}^{N_t}\sum_{\alpha\beta\gamma\delta}
\sigma_{\alpha\beta}(k)V_{\alpha\beta\gamma\delta}\sigma_{\gamma\delta}(k)\right ]
\nonumber \\
&\times& \exp\left [  -i\Delta t\sum_{\alpha\beta}(K_{\alpha\beta}+
\sum_{\gamma\delta}V_{\alpha\beta\gamma\delta}\sigma_{\gamma\delta}(k))\hat{\rho}_{\alpha\beta} \right ], 
\label{eq:tranamp}
\eea
where $t_f-t_i=N_t\Delta t$ (and $N_t\rightarrow\infty$), $V_{\alpha\beta\gamma\delta}$ are the matrix elements of the two-body interaction, $K_{\alpha\beta}$ the matrix elements of the kinetic energy (up to some simple renormalization), $\hat{\rho}_{\alpha\beta}=a^\dagger_\alpha a_\beta$ is bilinear in the creation and annihilation operators, and $\sigma_{\alpha\beta}(k)$ are the matrix elements of an auxiliary field. The stationary phase approximation for this evolution operator leads to a time dependent meanfield approximation, which however it is not uniquely defined \cite{qmps}. It has been shown that the one-body evolution operator (the last exponent in Eq. (\ref{eq:tranamp}) can be chosen to have an arbitrary form \cite{kerman} and that  this arbitrariness is resolved only when one goes beyond the stationary phase approximation and includes the quadratic fluctuations of the auxiliary field $\sigma_{\alpha\beta}(k)$. 

The fact that the stationary phase approximation is not uniquely defined is somewhat unsettling, and begs the question of whether the {\em best stationary phase approximation} can be introduced. Likely the answer is given within the TDDFT to be discussed below. At the same time, the fact that the non-uniqueness of the stationary phase approximation of the real-time functional integral approach is resolved only after the fluctuations of the auxiliary field are considered seems to suggest that the dynamics is truly governed by fluctuations and as such they always have to be accounted for. This aspect will be discussed in more detail below.

At the stationary phase approximation level however one can establish a quantization condition, which determines both the ground and the excited states. In Ref. \cite{qmps} it was shown that the resolvent operator for a many body system can be introduced as follows:
\beq
 G(E)=-i\int_0^\infty dt \mathrm{Tr} \exp(iEt-i\hat{H}t)
= -i\int_0^\infty dt A(t)\exp(iEt) =-i\int_0^\infty dt a(t)\exp[iEt+iS(t)], 
\label{eq:G(E)}
\eeq
where the trace $A(t)=\mathrm{Tr} \exp(-i\hat{H}t)$ of the evolution operator (\ref{eq:tranamp}) has been separated into an exponent of a (large) phase $S(t)$ and an (assumed) slowly varying function $a(t)$ of the time $t$. The trace is performed over the entire Hilbert space of the many-body system under consideration. In the stationary phase approximation the phase $S(t)$ is the classical action and the amplitude $a(t)$ is further determined by evaluating the quadratic fluctuations in the auxiliary field $\sigma_{\alpha\beta}(k)$. It is a rather straightforward exercise to show that by applying the stationary phase approximation to Eq. (\ref{eq:G(E)}) one obtains the quantization condition \cite{qmps}
\beq
 (2n+1)\pi  = i\int_0^{T(E)} dt 
\int dx \sum_\alpha\phi_\alpha^*(x,t)\frac{\partial}{\partial t}\phi_\alpha(x,t),
\label{eq:2pi}
\eeq
where the single particle wave functions satisfy the time-dependent meanfield equations
\beq
i\frac{\partial}{\partial t} \phi_\alpha(x,t) = h(x,t)\phi_\alpha(x,t)=
[K(x)+U(x,t)-\varepsilon_k(T)]\phi_\alpha(x,t),
\label{eq:h(t)}
\eeq
with the periodic time boundary conditions $\phi_\alpha(x,0)=\phi_\alpha(x,T)$. The auxiliary field $\sigma(x,t)$ is determined from the condition
\beq
\sigma(x,t)=\sum_\alpha|\phi_\alpha(x,t)|^2, \quad U(x,t) = \int dy V(x-y)\sigma(y,t) \label{eq:sigma}
\eeq
and the period $T(E)$ is determined from the equation
\beq
E=-\frac{\partial S(T)}{\partial T}=- \frac{\partial}{\partial T}
\left [\int_0^T dtdxdy V(x-y) \sigma(x,t)\sigma(y,t) -T\sum_k\varepsilon_k(T)\right ],
\label{eq:T(E)}
\eeq
(In Eq. (\ref{eq:2pi}) one should replace $2n+1$ by $2n$ if there are no {\em turning} points.) Even though here these equations have been written down in their simplest from, when the time-dependent meanfield is chosen to have a Hartree form, the formal structure of these equations remains basically the same if one chooses various other possibilities (Hartree-Fock or Hartree-Fock-Bogoliubov for example). The quantization condition (\ref{eq:2pi}) essentially amounts to requiring that the so-called Berry phase (obtained after removing the dynamical phase is quantized), see Refs. \cite{berry}. 

The stationary phase approximation of the real-time functional integral approach is well defined and straightforward to implement. However, the inclusion of the fluctuations of the auxiliary field amount to the evaluation of a functional integral of a complex exponential. If the amplitude of the fluctuations are small, one can expect that this integral could be evaluated numerically. It is still a largely unresolved question however of how to evaluate such an integral when the size of the fluctuating terms are large, though {\em some light of the end of the tunnel} is visible, see Refs.  \cite{qs,negele} and below.

\subsection{Time-dependent Density Functional Theory approach its limitations}

The second approach is the Density Functional Theory (DFT) \cite{HK}, extended to describe superfluid systems \cite{slda}. In the TDDFT extension of the approach, which has been studied for normal systems for quite some time \cite{td-dft,tddft}, one needs to extend the form the energy density functional in order to allow for currents, which are typically absent in the ground states. In many cases  requiring that the energy density functional to satisfy Galilean invariance allows to extend the DFT approach to a rather large class of  phenomena \cite{slda,todd}. As in the case of the stationary form of DFT, within TDDFT one can in principle evaluate exactly the  one-body observables, when the many-body system under consideration is subject to an arbitrary one-body  external potential \cite{td-dft,tddft}. The existence of such a formal result does provide us with an answer to a question raised  within the functional integral approach, namely: Could one define the best stationary phase approximation evolution operator? In order to define such an evolution operator one would have to introduce some criterion. The obvious choice in this regard appears to be an one-body evolution operator which would lead to a correct description of one-body observables. It is not clear however that with such a choice one can also ensure the fastest convergence of the formalism beyond the one-body level, which would be a more valuable requirement.  Within the TDDFT approach the problems discussed above concerning the choice of collective variables becomes superfluous and it is replaced instead by the problem of determination of periodic trajectories and determination of a solution of the quantization condition (\ref{eq:2pi}). 

The implementation of DFT in the case of nuclei faces a number of rather serious difficulties \cite{self}. As in the case of electrons, there is no theoretical recipe on how to construct a nuclear energy density functional.  Nuclei, unlike electrons, which reside in a somewhat well defined external potential created by the ions and for which a certain TDDFT formalism has been elaborated \cite{mdft}, are self-bound systems and there are difficulties with introducing a one-body density formalism. At this time it is not yet clear whether a viable solution exists, even though a number of suggestions has been put forward by various authors \cite{self}. Most researchers believe that this formal difficulty is not relevant in the study of medium and heavy nuclei, as the existence of single-particle excitation modes in nuclei is rather well established for more than half a century. A central tenet of DFT is that the correlation energy can be successfully incorporated into an exact energy density functional. In nuclei however the situation does not seem to be so clear. In particular, the gain in energy one obtains in constructing a rotationally invariant solution for open shell nuclei, which have intrinsic deformation, is a form of correlation energy and, in the opinion of most, see for example Refs. \cite{delaroche}, does not appear to be contained within any conceivable nuclear energy density functional. The same applies to other collective degrees of freedom, such as pairing \cite{duguet}, and in the case of relatively soft nuclei to the fluctuation of the shape parameters \cite{delaroche}. The only meaningful way to account for this type of correlation energy in case of nuclei appears to be either the introduction of a projection of the intrinsic many-body wave function (typically a GSD) or the the need to perform a GCM calculation (which often is reduced to the solution of a collective Schr\"{o}dinger equation) on top of the DFT calculation for the intrinsic wave function. In principle similar difficulties should arise in the case of extremely floppy molecules in chemistry \cite{mdft}. In the case of projection on good quantum numbers and of GCM calculations based on a DFT treatment of the intrinsic nuclear motion one needs to introduce non-diagonal in collective degrees of freedom expectation values for the internal energy, for which so far only prescriptions have been suggested \cite{duguet}.  

The issues raised above are just a forerunner of much deeper problems one encounters when applying the meanfield approach to LACM. The analog of the type of problems one faces in this case is the well known phenomenon of crossing of molecular terms. When two atoms approach each other they might undergo an electron excitation or not at some  separation. When such an internal excitation occurs, in which case an electron can be transferred from one atom to another and thus leading to formation of ions, the meanfield experienced by the ions/atoms changes. The rate of electron excitation depends very strongly on the ion/atom relative speed and it is often described as a Landau-Zener like transition. In order to describe the random character of such transitions and of their influence on the motion of nuclei a stochastic extension of the  meanfield approximation - the surface-hopping formalism - has been developed in chemistry by Tully \cite{tully} and this approach has been incorporated in modern versions of TDDFT. Another example discussed by Tully \cite{tully} is that of an atom impacting on a metal surface. Upon reflection such an atom will follow drastically different trajectories, depending on whether an electron-hole pair has been excited or not in the metal. If the metal is not excited, the atom will reflect elastically at an angle of reflection equal to the angle in incidence. However, if an electron-hole pair is excited within the metal surface, a noticeable amount of energy and linear momentum is being transferred to the metal and the atom will follow a trajectory at a much large angle of reflection than the incidence angle. Because of the existence of such processes the time honored Born-Oppenheimer approximation (in which the nuclear and electronic motion are adiabatically separated) is violated and corrections should be taken into account. As the work of Tully and others has demonstrated, these non-adiabatic corrections can be largely reduced to random transitions occurring mostly when various molecular terms cross. 

In the case of nuclear dynamics the equivalent of molecular terms is played by various potential energy surfaces. The case studied a long time ago by Negele and collaborators \cite{s32}, of the ``spontaneous fission'' of $^{32}S$,  is a great example of the need to modify the TDHF approach to nuclear dynamics and of the role of the potential energy surfaces crossing in such processes. By artificially increasing the proton electric charge these authors forced $^{32}S$ to split into two $^{16}O$ fragments. As these authors have shown, during the splitting  of the initial nucleus the system breaks the ground state symmetry of $^{32}S$  and develops a transitory octupole moment. This behavior can be interpreted as the influence of a diabolic point and the emergence of a gauge field located at a level crossing  \cite{berry}. Unlike the surface-hoping mechanism of Tully, there does not seem to be a need for a ``stochastic'' ingredient in order to allow a system to perform jumps from one potential energy surface to another, if one allows for the natural emergence of non-abelian gauge fields  \cite{berry,mead}. 

In nuclear dynamics it was recognized a long time ago that one needs to go beyond the meanfield approximation. In a TDHF approach, as in the TDDFT method, one can hope to describe correctly only one-body observables at most. In case of nuclear reactions between two heavy ions for example one would not be able to describe adequately the widths of the fragment mass distribution.  However, a long time ago Balian and V\'en\'eroni suggested a simple extension of the TDHF approximation in order to calculate second moments only of one-body observable \cite{balian}, and thus allowing us to overcome to some extent the restraints of a one-body formalism such as TDHF or TDDFT. The need to go beyond the simple TDHF framework has been argued in nuclear theory for quite sometime, see  Refs. \cite{shape,gupta,stoch}: in the role of the zero-point oscillations of the nuclear shape \cite{shape}, in the role of two-body collisions in affecting the meanfield behavior and leading to thermalization of the single-particle dynamics \cite{gupta}, and the need to extend in some ways the meanfield approach in order to allow for the fluctuations of the single-particle properties \cite{stoch}.  While apparently these extensions of the  meanfield approximation appear physically well founded and the agreement with experiment is most of the time satisfactory (partially due to the existence of an appropriate set of fitting parameters), the formal argumentation is somewhat lacking. 

\subsection{ Bringing the parts together }

On one hand the TDDFT is very appealing, as at least in principle allows an exact calculation of one-body observable. However, TDDFT does not seem to have an obvious extension if one would like to calculate many-body observables.  One the other hand, the functional integral method does not seem to be limited in principle, only if one would be able to really implement it in practice. And also, the functional integral approach appears to substantiate in the stationary phase approximation the existence of a  meanfield approach, even though this is not uniquely identified. On physical grounds we do not seem to need much of an argument to introduce some kind of surface-hopping within a TDDFT approach. A mere look at a Nilsson diagram will suffice to convince anyone that during a LACM a nucleus is bound to come across many level crossings. As it was argued in Refs. \cite{fsea} the number of such crossings is likely of the order of $A^{2/3}$ in a fissioning nucleus. The existence of so many level crossings will invalidate at once the use a single energy potential surface to describe LACM. While evolving through a sequence of shapes a nucleus will most likely perform jumps from one energy surface to the next in an essentially stochastic manner and there will be a very small chance that while trying to follow in reverse the sequence of shapes the nucleus would likely manage to return to its initial state. As it was mentioned above, each such level crossing, is not merely a source of surface-hopping, but also a source of a dynamically generated gauge field. Thus, the level crossings will lead to two rather different effects, both of them depending in a critical manner of the collective velocity. By enabling the surface-hopping in an essentially stochastic manner these will be the main source of irreversibility and of the generation of entropy. At the same time, unlike the potential forces, level crossings will generate non-abelian ``magnetic'' type of forces, which will force the system to go around a diabolical point, as it was observed in the case of ``fissioning'' $^{32}S$ \cite{s32}.  

The two {\it ab initio} approaches described above seem to come together rather naturally now. While within the strict functional approach one could not identify the best stationary phase approximation, on which to further build corrections due to fluctuations of the auxiliary field $\sigma_{\alpha\beta}$, the TDDFT formalism appears to suggest its existence. At the same time, the {\it missing} surface-hopping element in the TDDFT, which is clearly required by the physics of LACM and the need to compute many-body observables, appears rather naturally within the functional integral approach, which provides a very specific {\em complex} measure for the fluctuations of the  meanfield.  The fact that the measure of these fluctuations is complex is certainly an worrisome fact, as the practical implementation and the numerical convergence of such a scheme is far from evident. However, the recent progress in implementing the complex Langevin  method \cite{aarts} and the simple argument that in the case of fluctuations of vanishing amplitude one should not encounter any problems, serve as a very good encouragement that the successful implementation of such an approach is not out of question. There are however a number of formal questions to be addressed still. One of them is, what two-body interaction should be used to control the amplitude of the fluctuations, see Eq. (\ref{eq:tranamp}). While the functional integral approach appears to suggest that the matrix elements of the bare interaction should be used, the lessons we have learned from the implementation of the Boltzmann-Uhling-Uehlenbeck equation to describe nuclear kinetics \cite{gupta} and our {\it physical intuition} would suggest that these matrix elements are most likely strongly renormalized and this is an aspect which has not been clarified yet. Since Eq. (\ref{eq:tranamp}) will most likely be evaluated using a Monte-Carlo approach there is a need for a Metropolis importance sampling algorithm appropriate for this type of functional integrals.
 
\section{ Concluding remarks }

If successfully implemented, the strategy outlined above will allow us to obtain in principle an exact solution of the Schr\"{o}diger equation for a nuclear system in particular, since formally the representation of the real-time evolution operator (\ref{eq:tranamp}) is exact. This representation is equivalent to a Langevin type of extension of the TDDFT formalism, which can be implemented in a straightforward manner in practice. As recent progress suggest, the numerical implementation of the Langevin dynamics for many-body systems \cite{sq,aarts} does not appear to be a hopeless endeavor anymore \cite{aarts}. In this approach the transition amplitude from an initial state $\Psi_i$ to an arbitrary final state $\Psi_f$ is calculated from the expression:
\bea
\langle \Psi_f\left | \exp[-i\hat{H}(t_f-t_i)]\right |\Psi_i \rangle &=& 
\int \Pi_{k=1}^{N_t}\Pi_{\alpha\beta}\quad  d\sigma_{\alpha\beta}(k) 
\exp\left [ \frac{i\Delta t}{2}\sum_{k=1}^{N_t}\sum_{\alpha\beta\gamma\delta}
\sigma_{\alpha\beta}(k)V_{\alpha\beta\gamma\delta}\sigma_{\gamma\delta}(k)\right ]
\cal{U}[\sigma], 
\label{eq:stranamp}
\eea
where $\cal{U}[\sigma]$ is the one-body evolution operator for a specific spatio-temporal realization of the auxiliary field $\sigma_{\alpha\beta}(k)$, the last exponent in Eq.(\ref{eq:tranamp}). The {\em best} stationary phase approximation of this transition amplitude is given by the TDDFT evolution. This expression is similar in spirit to what one would see in a low intensity electron beam two-slit experiment, where one records one hit of the screen at a time and interference pattern appears only after adding up a large number of events. In order to obtain the full solution of the many-body problem, in particular a specific transition amplitude, one has to ``record'' many spatio-temporal realizations of the auxiliary filed $\sigma_{\alpha\beta}(k)$. If one is interested in a specific transition amplitude only, the optimal stationary phase approximation can be determined by adjusting it to the given initial and final states \cite{qmps,ripka}. 

A definite advantage of this approach is that it obviates the need to determine a set of collective variables. Various strategies devised so far in the literature to find such collective variables are based on an approximate solution of the time-dependent meanfield equations of motion such as Eqs. (\ref{eq:h(t)}) in the strict adiabatic limit. Adiabaticity is broken most prominently at level crossings and the need for surface-hoping has been advocated for a long time in literature \cite{tully,shape,gupta,stoch}. By solving Eqs. (\ref{eq:h(t)}) exactly, see Ref. \cite{bk,qmbnp}, the need to construct collective variables, potential energy surfaces and the corresponding mass tensor is obviated. Phenomena such as dissipation and fragment mass and energy distributions do not seem anymore out of reach for an {\it ab initio} inspired approach. Within an ATDHF approach in a five-dimensional collective space, which has been advocated by P. Moller and collaborators as a minimal one for fission dynamics \cite{pmoller}, one would need to generated a few million different configurations. And even when such a goal is achieved one would still obtain only a one-body description of the dynamics at best, and not arguably the most accurate. The generation of an ensemble of a few million realization of the auxiliary field $\sigma_{\alpha\beta}(k)$ on the other hand will allow for a significantly more physically complete description of the nuclear dynamics of a heavy nucleus.


Support is acknowledged from the DOE under grants DE-FG02-97ER41014 and DE-FC02-07ER41457. 


\end{document}